\documentclass{elsart}

\usepackage{epsfig}
\usepackage{amsmath}
\usepackage{latexsym}
\usepackage{exscale}

\renewcommand{\vec}[1]{\mbox{\boldmath{$#1$}}}

\begin{document}

\begin{frontmatter}

\title{On image reconstruction with the
       two-dimensional interpolating resistive readout structure
       of the Virtual-Pixel detector\thanksref{EU}}

\thanks[EU]{Work supported by the European Community
            (contract no. ERBFMGECT980104).}

\author[Siegen]{H. Wagner\corauthref{Autor}},
\author[Siegen]{A. Orthen},
\author[Siegen]{H.J. Besch},
\author[Siegen]{S. Martoiu},
\author[Trieste]{R.H. Menk},
\author[Siegen]{A.H. Walenta},
\author[Siegen]{U. Werthenbach}

\corauth[Autor]{Corresponding author.
Tel.: +49 271-740-3563;
fax:  +49 271-740-3533;
e-mail: wagner@alwa02.physik.uni-siegen.de.}

\address[Siegen]
  {Universit\"at Siegen, Fachbereich Physik,
   Emmy-Noether-Campus,
   Walter-Flex-Str. 3, 57068 Siegen, Germany}
\address[Trieste]
  {Sincrotrone Trieste, S.S. 14, km 163.5,
   Basovizza, 34012 Trieste, Italy}

\begin{abstract}
The two-dimensional interpolating readout concept of the
Virtual-Pixel detector (ViP detector) goes along with an enormous
reduction of electronic channels compared to pure pixel devices.
However, the special concept of the readout structure demands for
adequate position reconstruction methods. Theoretical
considerations reflecting the choice and the combination of linear
algorithms with emphasis on correct position reconstruction and
spatial resolution are presented. A subsequent two-dimensional
coordinate transformation further improves the image response of
the system. Measured images show how far the theoretical
predictions of the simulations can be verified and to what extent
they can be used to improve the position reconstruction.

\vspace{4mm}
\end{abstract}


\end{frontmatter}

\section{Introduction}

In general, position sensitive pixel detectors feature the
advantages of parallel and asynchronous readout which means i.e. a
proportional increase of the maximum incoming rate with the
sensitive detection area. However, these pure pixel devices suffer
from the enormous number of electronic channels leading either to
high costs and effort or essential size limitation of the
sensitive area. In contrast to that, interpolating readout
concepts which can cover large areas with a position resolution
comparable to pure pixel devices can also be carried out with
parallel and asynchronous readout. They combine the advantages of
pixel detectors with a reduction of electronics and costs at the
same time. For example the interpolating readout can be realised
by strips or pads
\cite{Bressan,Bachmann2002a,Cussonneau2002,Lewis1994}. We have
developed a truly two-dimensional interpolating readout concept
based on resistive charge division \cite{Besch}. Since this
concept requires to count single photons the signal of each event
has to be amplified, e.g. by gas amplification. For that purpose
e.g. micro pattern devices like GEM\footnote{GEM = {\textbf{g}}as
{\textbf{e}}lectron {\textbf{m}}ultiplier} \cite{Sauli},
micromegas\footnote{micromegas = \textbf{micro} \textbf{me}sh
\textbf{ga}seous \textbf{s}tructure} \cite{Giomataris1996} and
CAT\footnote{CAT = {\textbf{c}}ompteur {\textbf{\`{a}}}
{\textbf{t}}rous} \cite{Bartol} can be applied. We have combined
the readout structure with (optimised) MicroCAT gas gain
structures \cite{Sarvestani1998a,Orthen2003a} and a triple-GEM
configuration \cite{Orthen2003b}. In combination with the
triple-GEM configuration this readout concept has already proven
its possible application in biological and chemical diffraction
measurements \cite{Orthen2003c}. In principle this resistive
interpolating readout structure can also be applied with vacuum
devices like micro channel plates.

Beside the advantages of a good time resolution
\cite{Sarvestani2001} and high rate capability this
two-dimensional interpolation concept demands on the other hand
for a sophisticated choice of position reconstruction algorithms
and correction functions.

\section{Detector setup and working principle}

\begin{figure}
  \vspace{0mm}
  \begin{center}
    \includegraphics[clip,width=10cm]{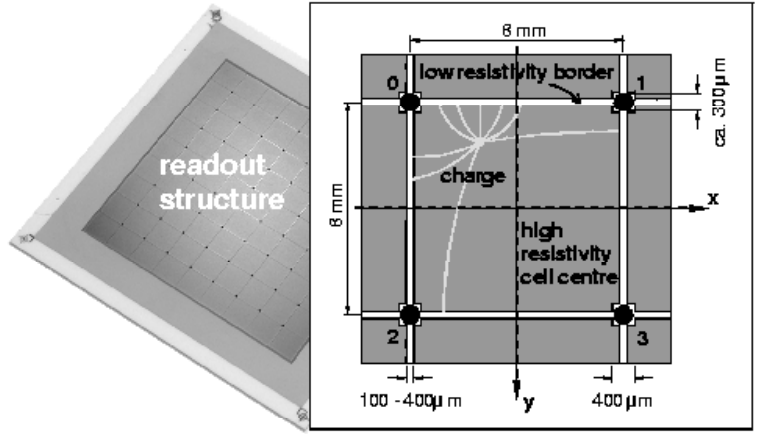}
 \end{center}
  \caption{Schematic of
           the two-dimensional interpolating
           resistive readout anode printed on a ceramic substrate.
           The gas gain structure (e.g. GEM or MicroCAT) is
           mounted above the
           anode.
           The structure is subdivided into $9\times 9$ cells,
           whereby the inner $7\times 7$ cells, corresponding
           to a sensitive area of $56\times 56\,\mathrm{mm^2}$, can be
           read out.
           A detailed magnification of one single cell
           of the interpolating readout structure on
           a ceramic substrate
           is illustrated on the right hand side.
           The charge, generated by a photon, is collected by readout
           nodes situated at the cell corners
           (represented by the black circles). The robust readout
           structure itself is absolutely insensitive against
           sparking or uncontrolled discharges.
           }
  \label{fig_readout_structure}
\end{figure}

Primary charge generated in the gas conversion region by incoming
photons (typical energy range: $5-25\,\mathrm{keV}$) via
photoelectric effect is led by an homogeneous electric drift field
towards the gas gain region where it is multiplied in an avalanche
process. This charge cluster (a so called event) hits the readout
plane, designed as a two-dimensional resistive structure, where
the sensitive area is subdivided into $7\times 7$ square, $8\times
8\,\mathrm{mm^2}$ sized cells (Fig. \ref{fig_readout_structure}).
Each cell is provided with two different surface resistivities in
order to obtain both an almost linear charge division behaviour in
both coordinates and a good electrical shielding to avoid too
large charge distributions over several cells. The high
resistivity centres of the cells are surrounded by low resistivity
narrow border strips (s. detail in Fig.
\ref{fig_readout_structure}). This leads to an almost
perpendicular projection of the charges in $x$- and $y$-direction
onto the low resistivity strips. The surface resistances of the
high resistivity cell centres and the low resistivity strips are
typically $R1=100$--$1000\,\mathrm{k\Omega}/\Box$ and
$R2=1$--$10\,\mathrm{k\Omega}/\Box$, respectively. Each cell is
read out at its four corners by low input impedance readout nodes.
The charges collected at the particular nodes are used to
determine the event position within the cell by means of a
suitable algorithm. All 64 readout nodes are equipped with charge
sensitive amplifiers and 12-bit FADCs which are linked to a
dedicated DAQ system which transfers the preprocessed data for
visualisation to a PC \cite{Martoiu}.

\section{The choice of reconstruction algorithms}

The first subsection describes the simulation tool which provides
the charge distribution within the readout structure. These
theoretical data will be used in the following to study the partly
unfavourable behaviour of different linear algorithms concerning
position reconstruction and spatial resolution. Afterwards, the
treated algorithms are combined to one more complex algorithm
which unifies the advantages of the individual algorithms.

A remark concerning the notation: The expressions for the
derivatives $\partial x/\partial u$ and $\partial_u x$ are used
synonymous.

\subsection{The diffusion simulation model}

Due to the surface resistances and the underlying distributed
capacitance the readout structure behaves like an integrating
$RC$-element resulting in a temporal broadening of the input
signal. The simulation of this dynamic charge diffusion process on
the readout structure has been realised by means of a numerical
computation programme, taking $4\times 4$-cells into account in
order to include charge flow into adjacent cells (boundary
effects). A very detailed derivation of the diffusion model and
its numerical solution can be found in Ref. \cite{Wagner2002a}. In
the diffusion model a space and time dependent driving current can
be impinged at every grid point. In the simulation, presented
here, we assumed a typical two-dimensional gaussian-like
transverse extension of the charge cloud of
$\sigma=200\,\mathrm{\mu m}$ (mainly caused by the transverse
diffusion of the primary electrons in the conversion gap).

The $4\times 4$-cell model provides $161\times 161$ grid points
which correspond to a homogeneous grid point spacing of $\Delta
x=\Delta y=200\,\mathrm{\mu m}$. The low resistivity border strips
have a width of $w=200\,\mathrm{\mu m}$, hence they are
represented by one single line of grid points. If not stated
differently the readout nodes are represented by one grid point
which corresponds to $200\times 200\,\mathrm{\mu m^2}$. Since the
readout nodes can be well approximated as ideal drains
\cite{Wagner2002a} the potential at the nodes is equal to zero at
all times.

This time-depending diffusion simulation is able to provide the
currents $I_n(t)$ and the charges $Q_n(t)$ at every particular
node $n$ for all times $t$. These simulated charges $Q_n(t)$ can
now be used to reconstruct the event position $(u,v)$ by means of
a suitable algorithm. Depending on the algorithm the reconstructed
position $(u,v)$ will differ up to a certain extent from the
(true) impact position $(x,y)$. Since this simulation model has
proven that its results are in good accordance to the measurements
\cite{Wagner2002a} it is used in the following to obtain the
relationship between the two spaces $(u,v)$ and $(x,y)$ to
optimise the applied reconstruction algorithms. For that reason
all further steps are directly based on the predictions of the
diffusion simulation.

\subsection{The linear reconstruction algorithms \label{Sec_linearAlgos}}

Only in the theoretical case of $R1/R2=\infty$ the charge division
in two dimensions is exactly linear. The required high rate
capability limits $R1$ upwards since large values of $R1$ lead to
stronger temporal charge diffusion and hence to a slower charge
collection at the readout nodes. On the other hand the parallel
resistive noise which determines the spatial resolution
significantly restricts $R2$ downwards. Therefore, only finite
resistivity ratios of $R1/R2$ will be realised in practice. For
that reason any linear reconstruction method produces approximate
positions only.

All our attempts introducing non-linear terms of the collected
charges $Q_n$ in a position reconstruction algorithm were not
successful under the boundary condition of correct reconstruction
at certain symmetry points of a cell. The discontinuity caused by
the low resistivity cell borders which have been introduced to
prevent the charge from leaving the cell and thus to improve the
high rate behaviour obviously complicate this attempt enormously.
Therefore these non-linear approaches are not further considered.

\begin{figure}
  \vspace{0mm}
  \begin{center}
   \includegraphics[clip,width=7.5cm]{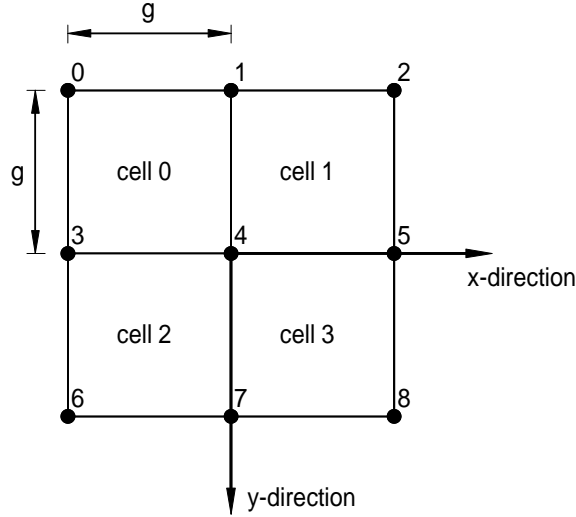}
  \end{center}
  \caption{This figure shows the node- and the cell-numeration. The origin of the
           coordinate system for the 6- and the 3-node algorithm is located
           at node 4. For the 4-node algorithm the origin is chosen in the
           centre of cell 1. The cell size
           is denoted with $g$.}
  \label{fig_node_indication}
\end{figure}

The simplest and most obvious linear reconstruction methods are
the so called 4-, 6-, and 3-node reconstruction algorithms,
whereby the node indications and the coordinate systems of the
particular algorithms are given in Fig. \ref{fig_node_indication}.
The correlation between the collected charges $Q_n$ at the
particular nodes $n$ and the reconstructed event positions $(u,v)$
can be realized for the
particular algorithms in the $x$- and $y$-direction as follows:\\
4-node algorithm (events somewhere in cell 1; $x,y\in
[-\frac{g}{2},\frac{g}{2}]$) :
\begin{align}\label{eq_4-node-algo}
    \begin{split}
     u_4& =\frac{g}{2}\cdot
          \frac{(Q_2+Q_5)-(Q_1+Q_4)}{Q_{xy4}}
          \quad\text{with}\quad Q_{xy4}=Q_1+Q_2+Q_4+Q_5\\[3mm]
     v_4& =\frac{g}{2}\cdot
          \frac{(Q_4+Q_5)-(Q_1+Q_2)}{Q_{xy4}}
     \end{split}
\end{align}
6-node algorithm (node 1 and 4 (node 4 and 5) carry maximum signal
for reconstruction in $x$-direction ($y$-direction);  $x \in
[0,\frac{g}{2}]$ and $y \in [-\frac{g}{2},0]$):
\begin{align}\label{eq_6-node-algo}
    \begin{split}
    u_6& =g\cdot
          \frac{(Q_2+Q_5)-(Q_0+Q_3)}{Q_{x6}}
          \quad\text{with}\quad Q_{x6}=Q_0+Q_1+Q_2+Q_3+Q_4+Q_5\\[3mm]
    v_6& =g\cdot
          \frac{(Q_7+Q_8)-(Q_1+Q_2)}{Q_{y6}}
          \quad\text{with}\quad Q_{y6}=Q_1+Q_2+Q_4+Q_5+Q_7+Q_8
    \end{split}
\end{align}
3-node algorithm (node 4 carries maximum signal; $x,y \in
[-\frac{g}{2},\frac{g}{2}]$):
\begin{align}\label{eq_3-node-algo}
    \begin{split}
     u_3& =g\cdot\frac{Q_5-Q_3}{Q_{x3}}
          \quad\text{with}\quad Q_{x3}=Q_3+Q_4+Q_5\\[3mm]
     v_3& =g\cdot\frac{Q_7-Q_1}{Q_{y3}}
          \quad\text{with}\quad Q_{y3}=Q_1+Q_4+Q_7
    \end{split}
\end{align}
The sum of the collected charges in $x$- and $y$-direction of the
particular algorithms are abbreviated with $Q_{xy4}$, $Q_{x6}$,
$Q_{y6}$, $Q_{x3}$ and $Q_{y3}$. The higher the ratio of $R1/R2$
the more pronounced is the cell like character and the electrical
screening among the cells. This leads to a treatment of the
readout structure consisting of strongly separated cells which
makes the application of the 4-node algorithm stressing this cell
structure most favourable. In practise the charge screening of the
cells is always finite. Therefore, the application of other
algorithms featuring symmetry planes at the cell border (6-node
algorithm) and the readout node (3-node algorithm) is sensible.

The most simplest algorithm imaginable is the 2-node algorithm.
But since this algorithm only takes two nodes into account it is
strongly affected by systematic effects (like cross-talk or
variations of the amplification of the preamplifiers). Therefore,
the 2-node algorithm, when applied in practise, is inferior to the
other algorithms and is not further considered.

\subsection{Reconstruction properties of the linear algorithms}

\subsubsection{Position reconstruction}

By means of the $4\times 4$-cell diffusion simulation it is
possible to investigate the reconstruction behaviour of the three
algorithms. A series of individual simulations has been carried
out impinging a short driving current successively at every grid
point (homogeneous grid point spacing $\Delta x=\Delta
y=200\,\mathrm{\mu m}$). All positions are calculated after the 25
nodes of the $4\times 4$-cells have collected almost $100\,\%$ of
the impinged charge. Altogether one obtains therefore $41\times
41$ possible positions within one single cell.

\begin{figure}
  \vspace{0mm}
  \begin{center}
    \includegraphics[clip,width=14cm]{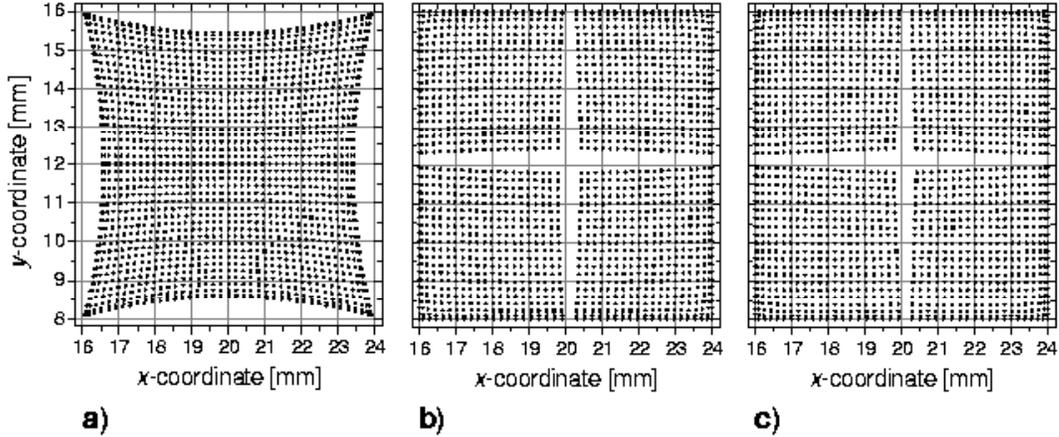}
  \end{center}
  \caption{
            Reconstructed positions
            within a $8\times 8\,\mathrm{mm^2}$ cell
            ($R1=100\,\mathrm{k\Omega}/\Box$ and
            $R2=2\,\mathrm{k\Omega}/\Box$) by means of
            \textbf{a)} the 4-node algorithm,
            \textbf{b)} the 6-node algorithm and
            \textbf{c)} the 3-node algorithm.
            }
  \label{fig_4er_6er_3er_Algo}
\end{figure}

In Fig. \ref{fig_4er_6er_3er_Algo} the reconstructed positions of
the 4-, 6- and 3-node algorithm are shown within a $8\times
8\,\mathrm{mm^2}$ cell for $R1/R2=50$.
The 4-node algorithm shows large distortions close to the cell
borders which means large deviations between the elements of the
two spaces $(x,y)$ and $(u,v)$, whereas this algorithm is nearly
distortion free in the centre of a cell. On the other hand both
the 6-node and the 3-node algorithm have their maximum distortions
in the cell centre. Close to the cell borders they are more
favourable for position reconstruction since their distortions are
much smaller compared to the 4-node algorithm. Obviously, the
reconstruction behaviour of the 6- and the 3-node algorithm is
quite similar, since they use comparable symmetry planes and
readout nodes.

For a homogeneous input illumination -- we impinged charges on
every grid point -- a perfect reconstruction algorithm would
result in a homogeneous output population density. Since none of
the introduced algorithms is a perfect reconstruction algorithm
(s. Fig. \ref{fig_4er_6er_3er_Algo}) variations in the population
density become obvious. Mathematically the correlation between the
true positions $(x(u,v),y(u,v))$ and the reconstructed positions
$(u,v)$ is given by a two-dimensional Jakobi-determinant $J(u,v)$
(coordinate transformation), which correlates the two population
densities $B(x,y)$ and $\widetilde{B}(u,v)$:
\begin{align}
    \int dx\,dy\,B(x,y)&=\int du\,dv\, J(u,v)\cdot
    B\big(x(u,v),y(u,v)\big)=\int du\,dv\, \widetilde{B}(u,v)
\end{align}
Therfore, we obtain
\begin{align}\label{eq_population_density}
\begin{split}
    \widetilde{B}(u,v)&=J(u,v)\cdot
    B\big(x(u,v),y(u,v)\big)=
        \left|
        \begin{array}{cc}
            \displaystyle \frac{\partial x(u,v)}{\partial u} & \displaystyle \frac{\partial y(u,v)}{\partial u}\\[4mm]
            \displaystyle \frac{\partial x(u,v)}{\partial v} & \displaystyle \frac{\partial y(u,v)}{\partial v}
        \end{array}
     \right|
 \cdot B(x,y)
 \end{split}
\end{align}
with the Jakobi-determinant $J$:
\begin{align}\label{eq_JakobiDeterminant}
 \begin{split}
   J &= \frac{\partial x}{\partial u}\frac{\partial y}{\partial v}
       -\frac{\partial x}{\partial v}\frac{\partial y}{\partial u}
      = \frac{1}{\displaystyle\frac{\partial u}{\partial x}\displaystyle\frac{\partial v}{\partial y}
                -\displaystyle\frac{\partial u}{\partial y}\displaystyle\frac{\partial v}{\partial x}
                }\,\,,
 \end{split}
\end{align}
whereby the second relationship for $J$ can be derived by the
back-trans\-formation $B(x,y)=\widetilde{J}(x,y)\cdot
\widetilde{B}(u(x,y),v(x,y))$. The derivatives with respect to the
true positions $\partial_x$ and $\partial_y$ are easier to
calculate in a numerical fashion since the positions $(x,y)$ are
placed on an equidistant lattice in the case of a homogeneous
illumination. Therefore the second relationship for $J$ from Eq.
(\ref{eq_JakobiDeterminant}) is used for further numerical
calculations.

\begin{figure}
  \vspace{0mm}
  \begin{center}
    \includegraphics[clip,width=7.5cm]{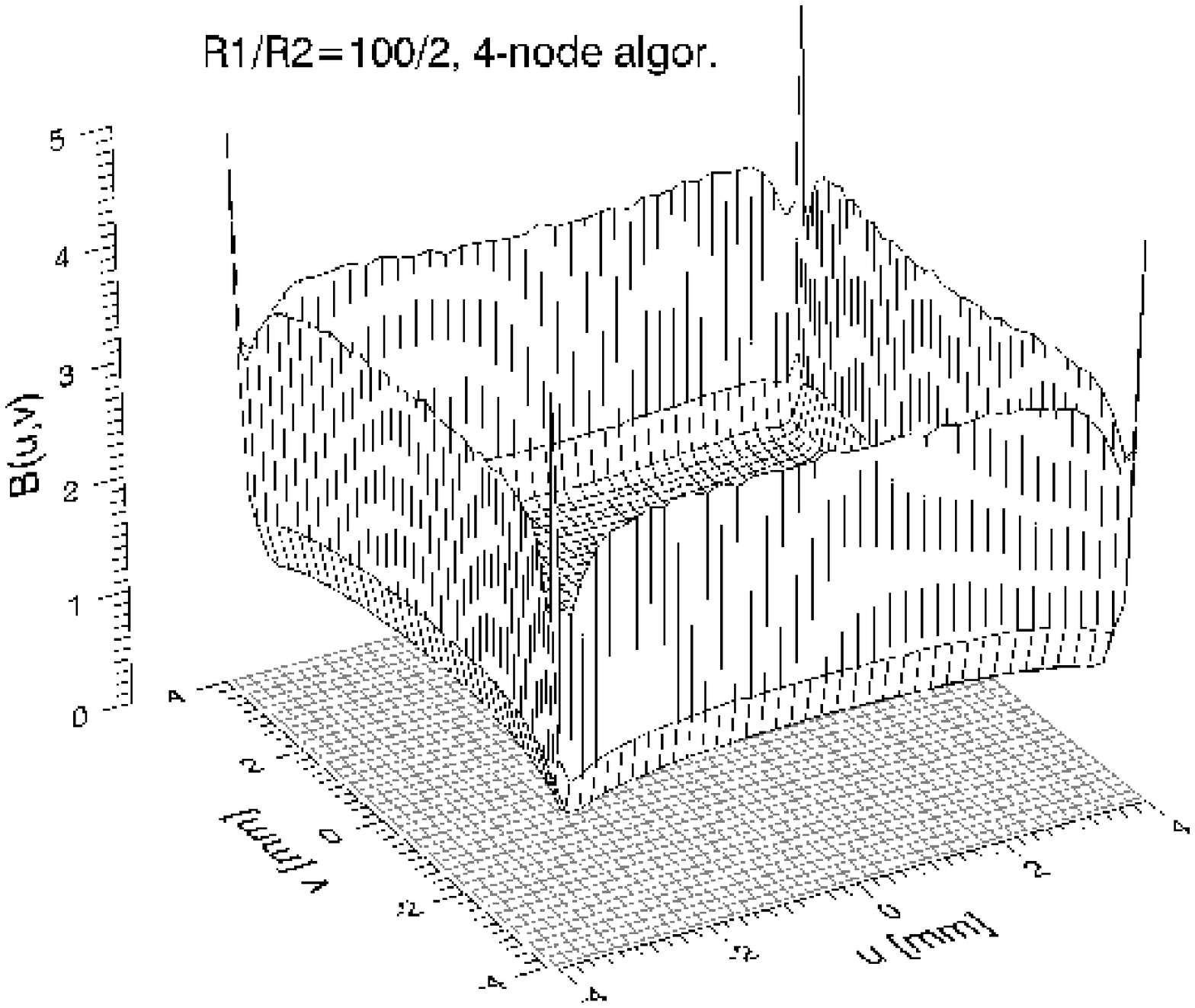}
 \end{center}
  \caption{
            Population density $\widetilde{B}(u,v)$ plotted for the 4-node algorithm
            for $R1=100\,\mathrm{k\Omega}/\Box$ and
            $R2=2\,\mathrm{k\Omega}/\Box$ obtained by
            Eq. (\ref{eq_population_density}) with a homogeneous
            illumination $B(x,y)=1$.
          }
 \label{fig_PopDensity4er}
\end{figure}

A homogeneous illumination like in this simulation means $B(x,y) =
\text{const.}$ which is set in the following (arbitrarily) to
$B(x,y) = 1$. This leads to $\widetilde{B}(u,v)=J(u,v)$ (Eq.
(\ref{eq_population_density})). In Fig. \ref{fig_4er_6er_3er_Algo}
the relationship between the true positions $(x,y)$ and the
reconstructed positions $(u,v)$ is shown for quantised distances.
It is obvious that $J$ is unequal to one which directly leads to
an inhomogeneous population density $\widetilde{B}(u,v)$. This is
also visible by the variation of the density of the black points
in Fig. \ref{fig_4er_6er_3er_Algo}. A higher density of points
$(u,v)$ is equivalent to a higher population density
$\widetilde{B}(u,v)$. In Fig. \ref{fig_PopDensity4er} the
population density $\widetilde{B}(u,v)$ for the 4-node algorithm
is shown as an example.
A perfect algorithm which connects the reconstructed positions
with the true positions $x(u,v)$ and $y(u,v)$ by $x(u)=u$ and
$y(v)=v$ would lead to a population density
$\widetilde{B}(u,v)=B(x,y)$.

Unfortunately, it is a non-trivial task to find a suitable
transformation between the two spaces $(x,y)$ and $(u,v)$ since
the functions $x(u,v)$ and $y(u,v)$ are explicitly needed.
Starting with one of the linear algorithms (e.g. the 4-node
algorithm) one will end up in major problems: Since all the data
were calculated numerically with a finite grid spacing (e.g.
$200\,\mathrm{\mu m}$) one has to approximate the functions
$x(u,v)$ and $y(u,v)$ e.g. by polynomials or Fourier series to get
an analytical approach. Thereby, it showed up that one has to take
too many orders into account to decrease the deviation between
positions reconstructed by the simulation and the polynomials. But
even if there exists a satisfying solution of an analytical
expression for $x(u,v)$ and $y(u,v)$ another much more serious
problem arises. One has to take carefully the spatial resolution
into account. As it will be seen later on, the transformation for
$x(u,v)$ and $y(u,v)$, which will correct the reconstructed
positions $(u,v)$, will dramatically deteriorate the spatial
resolution at the positions where larger corrections have to be
applied. This will be for example the case at the cell borders
when using the 4-node algorithm where we have larger population
densities (s. Fig. \ref{fig_PopDensity4er}) and therefore larger
position corrections.

Also the access via a direct measurement is closed since an
extensive mechanical scan has to be done; e.g. by means of a fine
needle impinging a certain amount of charge onto the readout
structure. At least one has to scan -- depending on the desired
accuracy -- a few thousand positions which will lead to an
enormous effort. Besides, the true positions $(x,y)$ are hard to
determine due to the finite mechanical accuracy of the setup.

\subsubsection{Spatial resolution \label{Sec_ConsiderationSpatialRes}}

In most position sensitive detector systems the spatial resolution
represents an important parameter which has to be optimised
depending on the desired application. Since in interpolating
systems like in our case the reconstruction algorithms themselves
affect the spatial resolution additionally (besides e.g. the
signal-to-noise ratio (SNR) or the pixel-size for pure pixel
detectors), the influence of the algorithms has to be analysed
separately in this respect.

The absolute values of the spatial resolution of the different
algorithms are determined by the SNR. Since the spatial resolution
of all algorithms is $\propto \text{SNR}^{-1}$ (s. b.) the
absolute noise resistance $R$ is not explicitly needed when the
various algorithms are compared to each other. In the following we
treat the parallel noise of the readout structure as the dominant
noise source, which mainly determines the spatial resolution.
Every readout node collecting the charge $Q_i$ is subjected to
Nyquist-Johnson noise
\begin{align}\label{eq_NyquistJohnsonNoise}
   \Delta Q(t) =
   \sqrt{
   \frac{4\,k_B T \cdot t}{R}
        }\,,
\end{align}
with $t$ denoting the charge integration time. Thereby, the
collected charge including the noise contribution is normally
distributed around $Q_i(t)$. For the total spatial resolution the
electronic noise would also have to be considered. It has
deliberately not been included in the calculations to leave the
discussion independent from the electronics used. For most
practical cases it will be in the same order of magnitude or
smaller than the noise contribution, presented here.

\begin{figure}
  \vspace{0mm}
  \begin{center}
   \includegraphics[clip,width=7.5cm]{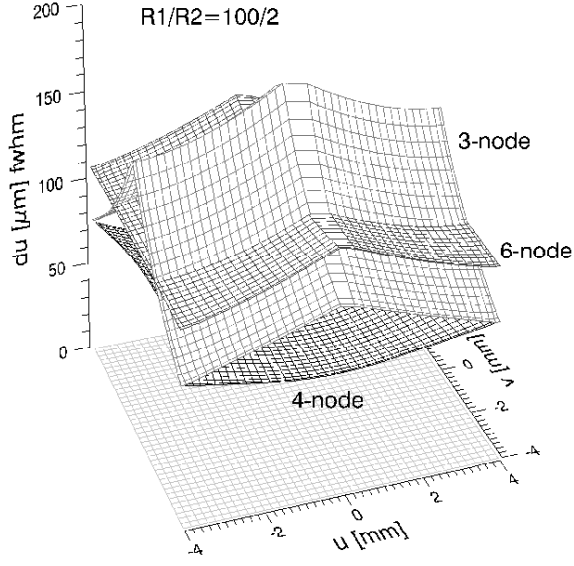}
 \end{center}
 \caption{  Spatial resolution $du=2.355\cdot\sigma_u$ (fwhm)
            for the 4-, 6- and 3-node algorithm
            as a function of the
            reconstructed positions $(u_n,v_n)$ for
            $R1=100\,k\Omega/\Box$ and $R2=2\,k\Omega/\Box$ after
            the complete charge has been integrated
            ($t=500\,\mathrm{ns}$).
            The driving current contains a charge equivalent of about $1.7\cdot
            10^6\,\mathrm{e^-}$.
            The parallel noise
            resistance of the readout structure has been
            numerically calculated to
            $R\approx 13.6\,\mathrm{k\Omega}$.}
 \label{fig_4_6_3SpatialResUncorrected}
\end{figure}

Taking into account the particular algorithms (Eqs.
(\ref{eq_4-node-algo})--(\ref{eq_3-node-algo})) and assuming a
constant gaussian charge noise $\Delta Q$ (Eq.
(\ref{eq_NyquistJohnsonNoise})) error propagation leads to the
spatial resolution in $u$-direction:
\begin{align}\label{eq_error_propagation}
 \sigma_u(u,t) =
   \left(
     \sum_i
     \left(
  \frac{\partial u}{\partial Q_i(x,y,t)}\,\Delta Q(t)
   \right)^2
  \right)^{1/2}\,,
\end{align}
where $(x,y)$ and $t$ indicate the true event position (impact
position) upon the readout plane and the charge integration time,
respectively. The calculation in $v$-direction behaves
correspondingly. The sum runs over all participating nodes of the
particular algorithms. Since some charges $Q_i$ appear both in the
$u$- and the $v$-coordinate the two coordinates are correlated.
Therefore, the noise causes a spatial smearing which can be
expressed by a two-dimensional gaussian distribution with
$\varrho$ describing the strength of the correlation (e.g.
\cite{Brandt}):
\begin{align}\label{eq_2DGaussianDistribution}
    \begin{split}
    E(x,y) &=
     \frac{1}{2\pi\,\sigma_u\sigma_v\sqrt{1-\varrho^2}}\cdot\\
     &\quad\exp
       \bigg\{
             -\frac{1}{2}\cdot\frac{1}{1-\varrho^2}
             \bigg(
                  \frac{(x-u)^2}{\sigma_u^2}+\frac{(y-v)^2}{\sigma_v^2}-
                  \frac{2\varrho(x-u)(y-v)}{\sigma_u\sigma_v}
             \bigg)
       \bigg\}
     \end{split}
\end{align}
For the 4-, 6- and 3-node algorithms one obtains for the spatial
uncertainty (corresponding coordinate systems s. Fig.
\ref{fig_node_indication}):

4-node algorithm:
\begin{align}\label{eq_spatial_resolution_4-node}
    \begin{split}
        \sigma_{u4}(u_4(t),t)
                 &= \sqrt{4u_4(t)^2+g^2} \cdot
                    \frac{\Delta Q(t)}{Q_{xy4}(x,y,t)}= f_{u4}(u_4(t),g)\cdot(\text{SNR}_{xy4})^{-1}\\
         \sigma_{v4}(v_4(t),t)
                 &= \sqrt{4v_4(t)^2+g^2} \cdot
                    \frac{\Delta Q(t)}{Q_{xy4}(x,y,t)}=
                    f_{v4}(v_4(t),g)\cdot(\text{SNR}_{xy4})^{-1}\\
         \text{cov}(u_4,v_4,t)
                 &= 4 u_4v_4
                    \bigg(
                            \frac{\Delta Q}{Q_{xy4}}
                    \bigg)^2 = 4 u_4v_4 \cdot(\text{SNR}_{xy4})^{-2}\\
        \varrho(u_4,v_4,t) &= \frac{4 u_4v_4}{\sqrt{(4u_4^2+g^2)(4v_4^2+g^2)}}
    \end{split}
\end{align}
6-node algorithm:
\begin{align}\label{eq_spatial_resolution_6-node}
    \begin{split}
         \sigma_{u6}(u_6(t),t)
                  &= \sqrt{6u_6(t)^2+4g^2} \cdot
                  \frac{\Delta Q(t)}{Q_{x6}(x,y,t)}=f_{u6}(u_6(t),g)\cdot(\text{SNR}_{x6})^{-1}\\
         \sigma_{v6}(v_6(t),t)
                  &= \sqrt{6v_6(t)^2+4g^2} \cdot
                  \frac{\Delta
                  Q(t)}{Q_{y6}(x,y,t)}=f_{v6}(v_6(t),g)\cdot(\text{SNR}_{y6})^{-1}\\
         \text{cov}(u_6,v_6,t)
                 &= \big(
                            4 u_6v_6 +2g(u_6-v_6)-g^2
                    \big)\,
                    \frac{(\Delta Q)^2}{Q_{x6}\,Q_{y6}}\\
         \varrho(u_6,v_6,t) &= \frac{4 u_6v_6 +2g(u_6-v_6)-g^2}{\sqrt{(6u_6^2+4g^2)(6v_6^2+4g^2)}}
    \end{split}
\end{align}
3-node algorithm:
\begin{align}\label{eq_spatial_resolution_3-node}
    \begin{split}
         \sigma_{u3}(u_3(t),t)
                  &= \sqrt{3u_3(t)^2+2g^2} \cdot
                  \frac{\Delta Q(t)}{Q_{x3}(x,y,t)}=f_{u3}(u_3(t),g)\cdot(\text{SNR}_{x3})^{-1}\\
         \sigma_{v3}(v_3(t),t)
                  &= \sqrt{3v_3(t)^2+2g^2} \cdot
                  \frac{\Delta
                  Q(t)}{Q_{y3}(x,y,t)}=f_{v3}(v_3(t),g)\cdot(\text{SNR}_{y3})^{-1}\\
         \text{cov}(u_3,v_3,t) &= u_3v_3\,
                    \frac{(\Delta Q)^2}{Q_{x3}\,Q_{y3}}\\
         \varrho(u_3,v_3,t) &= \frac{u_3v_3}{\sqrt{(3u_3^2+2g^2)(3v_3^2+2g^2)}}
    \end{split}
\end{align}
The cell size is denoted by $g$. Again the reconstructed positions
are indicated by $(u(x,y),v(x,y))$ together with a number as a
subscript for the particular algorithm. The terms of the spatial
resolutions can be separated into a geometrical factor $f$ and a
factor consisting of the SNR. Fig.
\ref{fig_4_6_3SpatialResUncorrected} shows schematically the
results of the spatial resolutions $du=2.355\cdot\sigma_u$ (fwhm)
as a function of the distorted reconstructed position $(u_n,v_n)$
for all three algorithms within a single $8\times
8\,\mathrm{mm^2}$-cell taking both the geometrical factor $f$ and
the SNR-factor into account.
Obviously, the 4-node algorithm shows up to be the most favourable
algorithm since it provides the best spatial resolution. Since the
3-node algorithm collects in the cell centre only about half of
the impinged charge the SNR-factor decreases and therefore the
spatial resolution deteriorates in comparison to the other
algorithms. Only at the nodes the spatial resolution for the 3-
and the 4-node algorithm becomes similar, since the geometric
factors $f=\sqrt{2}\,g$ (cf. Eqs.
(\ref{eq_spatial_resolution_4-node}) and
(\ref{eq_spatial_resolution_3-node})) and the SNR-factors are the
same.

\begin{figure}
  \vspace{0mm}
  \begin{center}
    \includegraphics[clip,width=7.5cm]{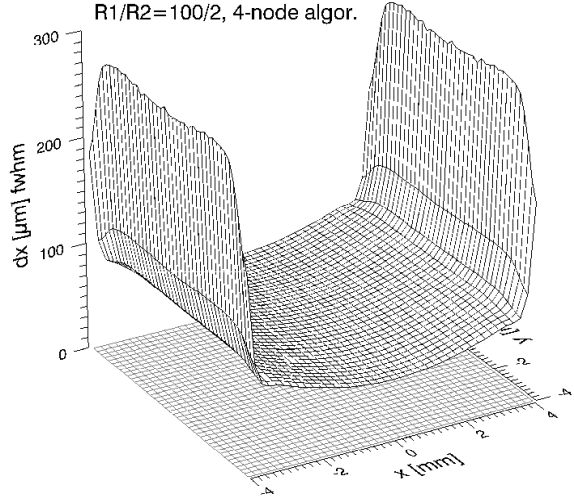}
 \end{center}
  \caption{
            Spatial resolution $dx=2.355\cdot\sigma_x$ (fwhm)
            for the 4-node algorithm
            as a function of the
            true positions $(x,y)$ for
            $R1=100\,k\Omega/\Box$ and $R2=2\,k\Omega/\Box$.
            The graph is plotted for the
            same parameters like in Fig.
            \ref{fig_4_6_3SpatialResUncorrected}. Since both the
            6- and the 3-node algorithm show quite the same
            behaviour close to the cell borders only the 4-node
            algorithm is plotted exemplarily.
         }
 \label{fig_4_6_3SpatialResCorrected}
\end{figure}

Actually, the spatial resolution in Fig.
\ref{fig_4_6_3SpatialResUncorrected} is plotted as a function of
the reconstructed positions $(u,v)$ and not of the true positions
$(x,y)$. If one corrects the reconstructed positions by means of a
coordinate transformation which transforms $(u,v)\rightarrow(x,y)$
the spatial resolution of the corrected positions $\sigma_x$ can
be estimated by:
\begin{align}\label{eq_SpatialResTrafo}
\begin{split}
    \big(\sigma_x(x,y)\big)^2 &= \frac{1}{
                \bigg(
                \displaystyle\frac{\partial u}{\partial x}\displaystyle\frac{\partial v}{\partial y}
                -\displaystyle\frac{\partial u}{\partial y}\displaystyle\frac{\partial v}{\partial x}
                \bigg)^2}\,\,\cdot\\[3mm]
    &\quad\Bigg\{
        \bigg(\frac{\partial v}{\partial y}\bigg)^2 (\sigma_u)^2 +
        \bigg(\frac{\partial u}{\partial y}\bigg)^2 (\sigma_v)^2
        -2\,\frac{\partial v}{\partial y}\frac{\partial u}{\partial y}
        \, \text{cov}(u,v)
    \Bigg\}
\end{split}
\end{align}
Eq. (\ref{eq_SpatialResTrafo}) determines the spatial resolution
$\sigma_x$ of the corrected positions as a function of the
covariance, the spatial resolutions $\sigma_u$ and $\sigma_v$ of
the applied algorithm (cf. Eqs.
(\ref{eq_spatial_resolution_4-node})--(\ref{eq_spatial_resolution_3-node}))
and the derivatives between the coordinates $(u,v)$ and $(x,y)$.
In this equation all appearing $(u,v)$ should be treated as
$(u(x,y),v(x,y))$. Fig. \ref{fig_4_6_3SpatialResCorrected} shows
exemplarily the numerically calculated spatial resolution
$dx=2.355\cdot\sigma_x$ (fwhm) obtained by Eq.
(\ref{eq_SpatialResTrafo}) as a function of the true positions
$(x,y)$ for the 4-node algorithm.
Since at the cell borders where the population density of all
algorithms is relatively high (s. Fig. \ref{fig_4er_6er_3er_Algo})
larger corrections have to be applied the spatial resolution
deteriorates by almost a factor of $3$--$4$ for all algorithms.

\subsection{Introduction of the optimised 463-node algorithm \label {Sec_463-node algorithm}}

In the following we present a possible solution which takes care
of both the correct position reconstruction and the spatial
resolution, by combining the 4-, 6- and 3-node algorithms to one
463-node algorithm.

All algorithms converting a homogeneous population density
$B(x,y)=1$ into an inhomogeneous population density
$\widetilde{B}(u,v)\neq 1$ which is equivalent to a wrong position
reconstruction are not favourable. Even for the case of a later
position correction these algorithms will always suffer from a bad
spatial resolution at the positions where larger position
corrections have to be applied (cf. end of Sec.
\ref{Sec_ConsiderationSpatialRes}). Therefore, it would be much
more advantageous to find an algorithm which is capable to
reconstruct almost the correct positions which require only small
subsequent position corrections and which therefore only weakly
influence the spatial resolution. The idea of combining the 4-, 6-
and 3-node algorithm to the 463-node algorithm shows up to be a
possible solution fulfilling almost the requirements mentioned
above.

\begin{figure}
  \vspace{0mm}
  \begin{center}
  \includegraphics[clip,width=7.5cm]{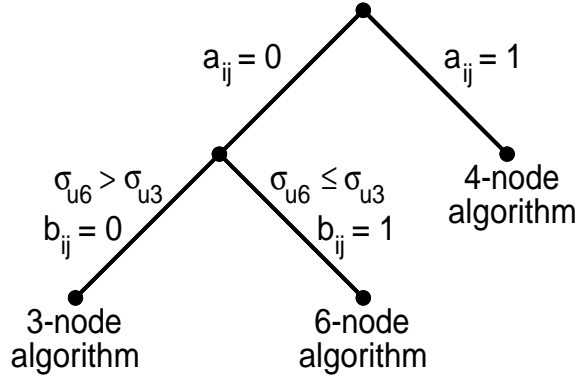}
 \end{center}
  \caption{
            Choice of the mixing matrices $A_x$ and $B_x$. The
            first bifurcation mixes between the 4-node and either
            the 6- or the 3-node algorithm. The criterion for $a_{ij}\in[0,1]$
            is an optimised position reconstruction which will be described in the text.
            The second bifurcation distinguishes between the 6- and the 3-node
            algorithm. The criterion for $b_{ij}\in \{0,1\}$ is given by the spatial
            resolution $\sigma_{u6}$ and $\sigma_{u3}$.
            }
 \label{fig_choice463er}
\end{figure}

Based on the results of the simulation model giving a relationship
between $(u,v)$ and $(x,y)$ the combination of the algorithms in
$x$-direction is done by mixing matrices $A_x=((a_{ij})_x)$ and
$B_x=((b_{ij})_x)$, defined for every grid point in the $8\times
8\,\mathrm{mm^2}$ cell. The elements $a_{ij}\in [0,1]$ mix the
4-node algorithm continuously with either the 6-node or the 3-node
algorithm, whereby the elements of the second mixing matrix
$b_{ij}\in \{0,1\}$ consisting only of the values 0 or 1 choose
either the 6-node ($b_{ij}=1$) or the 3-node ($b_{ij}=0$)
algorithm. The following equation defines the complete 463-node
algorithm, whereby $a_x=a_{ij}$ and $b_x=b_{ij}$ are those
elements of the matrices corresponding to the true position
$(x,y)$. The origin of the 4-node algorithm is now shifted to
readout node 4 (cf. Fig. \ref{fig_node_indication}):
\begin{align}\label{eq_463-node-algo}
\begin{split}
    u_{463}&= a_x u_4 + (1-a_x)
               \big\{
                  b_x u_6 + (1-b_x)u_3
               \big\}\\[2mm]
   &= a_x\,\frac{g}{2}\cdot
        \bigg\{
              \frac{(Q_2+Q_5)-(Q_1+Q_4)}{Q_{xy4}}+1
        \bigg\} +\\[2mm]
   &\quad\quad
        (1-a_x)\,g\cdot\bigg\{
                        b_x \cdot
                         \frac{(Q_2+Q_5)-(Q_0+Q_3)}{Q_{x6}}+
                        (1-b_x)\cdot
                        \frac{Q_5-Q_3}{Q_{x3}}
                \bigg\}
\end{split}
\end{align}
For symmetry reasons the relationship between the mixing matrices
in $x$- and $y$-direction is simply: $A_y = A_x^T$ and $B_y =
B_x^T$. The procedure for the combination of $A_x$ and $B_x$ is
shown in the schematic diagram in Fig. \ref{fig_choice463er}.
In the following the derivation of the matrices $A_x$ and $B_x$ is
discussed for the $x$-direction. The $y$-direction behaves
correspondingly.

The spatial resolution $\sigma_{u6}$ and $\sigma_{u3}$ of the 6-
and the 3- node algorithm is the criterion for the matrix $B_x$
with $b_{ij}\in \{0,1\}$ (cf. Eqs.
(\ref{eq_spatial_resolution_6-node}) and
(\ref{eq_spatial_resolution_3-node})). If $\sigma_{u6}\le
\sigma_{u3}$ the 6-node algorithm is used ($b_{ij}=1$); if
$\sigma_{u6} > \sigma_{u3}$ the 3-node algorithm is used
($b_{ij}=0$). Since the deviations between the reconstructed
positions of the 6-node and the 3-node algorithm are small (s.
Fig. \ref{fig_4er_6er_3er_Algo}) also the influence of the points
of discontinuity ($b_{ij}=0\,\Leftrightarrow\,b_{ij}=1$) is
expected to be negligible.

The criterion for the combination between the 4-node and the 6- or
the 3- node algorithm is an optimised position reconstruction
which means a minimum deviation $\Delta$ between $u$ and $x$.
Together with $u_{6,3}= b_x u_6 + (1-b_x)u_3$ we get as a
criterion:
\begin{align}\label{eq_criterion_a}
    \min \Delta =\min \big|x-u_{463}\big|^2 =
    \min\bigg\{
           \big(x-a_xu_4-(1-a_x)u_{6,3} \big)^2
    \bigg\} \,,
\end{align}
whereby again $a_x=a_{ij}$ and $b_x=b_{ij}$. The minimum condition
leads to $\partial \Delta/\partial a_x=0$. Since $\partial^2
\Delta/\partial a_x^2\ge 0$ the criterion for a minimum is
fulfilled. As a result we obtain for the matrix elements $a_x$:
\begin{align}\label{eq_solution_matrix_a}
    a_x=\frac{x-u_{6,3}}{u_4-u_{6,3}} \quad
    \begin{cases}
        \text{if } a_x>1 \text{ then } a_x \text{ is set to 1}\\
        \text{if } a_x<0 \text{ then } a_x \text{ is set to 0}
    \end{cases}
\end{align}
The constraint
\begin{align}\label{eq_constraint_Matrix_a}
    0\le a_x \le 1 \,,
\end{align}
has to be set because for some positions mixing factors $|a_x|\gg
1$ would end up in additional distortions due to noise or
systematic effects (cf. Sec. \ref{Sec_NonLinCorrections}).
Consequently, this constraint leads to the fact that some
positions $(u,v)$ are still not reconstructed correctly.

\begin{figure}
  \vspace{0mm}
  \begin{center}
  \includegraphics[clip,width=7.5cm]{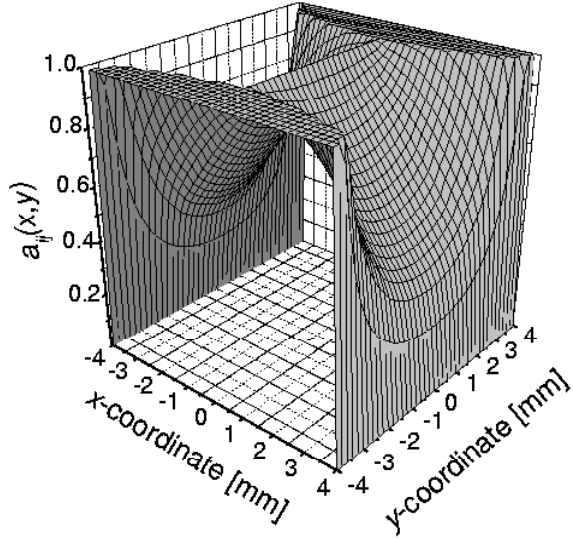}
 \end{center}
  \caption{
            Matrix $A_x$ with $a_x\in [0,1]$ calculated with Eq. (\ref{eq_solution_matrix_a}) for
            $R1=100\,k\Omega/\Box$ and $R2=2\,k\Omega/\Box$
            ($a_{ij}=1$ corresponds to
            the full usage of the 4-node algorithm, $a_{ij}=0$
            corresponds to
            the pure use of either the 6- or the 3-node algorithm).
        }
 \label{fig_matrix_a}
\end{figure}

\begin{figure}
  \vspace{0mm}
  \begin{center}
   \includegraphics[clip,width=7.5cm]{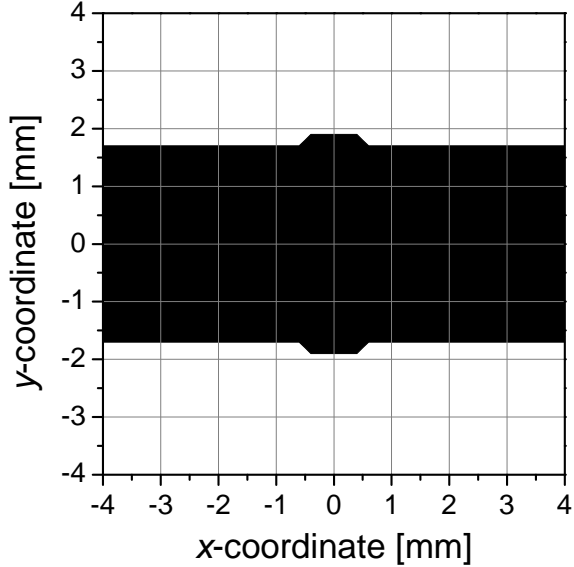}
 \end{center}
  \caption{
            Matrix $B_x$ with $b_x\in \{0,1\}$ plotted for
            $R1=100\,k\Omega/\Box$ and $R2=2\,k\Omega/\Box$ ($b_{ij}=1$
            (black area) corresponds to
            the full application of the 6-node algorithm and $b_{ij}=0$
            (white area) corresponds to
            the full application of the 3-node algorithm).
            }
 \label{fig_matrix_b}
\end{figure}

\begin{figure}
  \vspace{0mm}
  \begin{center}
    \includegraphics[clip,width=7.5cm]{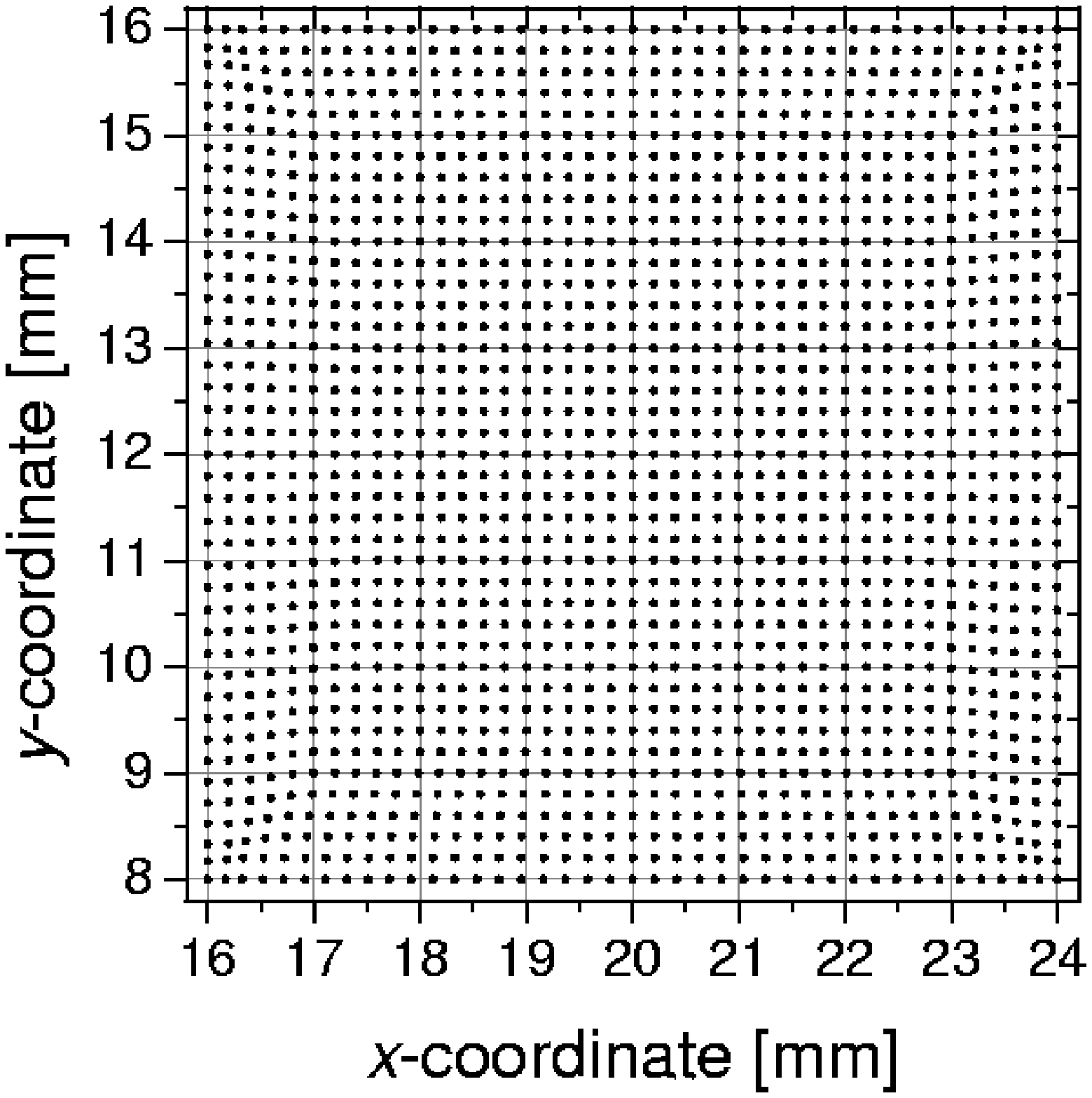}
 \end{center}
  \caption{
            Reconstructed positions
            by means of the 463-node algorithm
            within a $8\times 8\,\mathrm{mm^2}$ cell
            ($R1=100\,\mathrm{k\Omega}/\Box$ and
            $R2=2\,\mathrm{k\Omega}/\Box$). The corresponding
            mixing matrices $A_x$ and $B_x$ are shown in Figs.
            \ref{fig_matrix_a} and \ref{fig_matrix_b},
            respectively.
           }
 \label{fig_463node-algo}
\end{figure}

\begin{figure}
  \vspace{0mm}
  \begin{center}
  \includegraphics[clip,width=7.5cm]{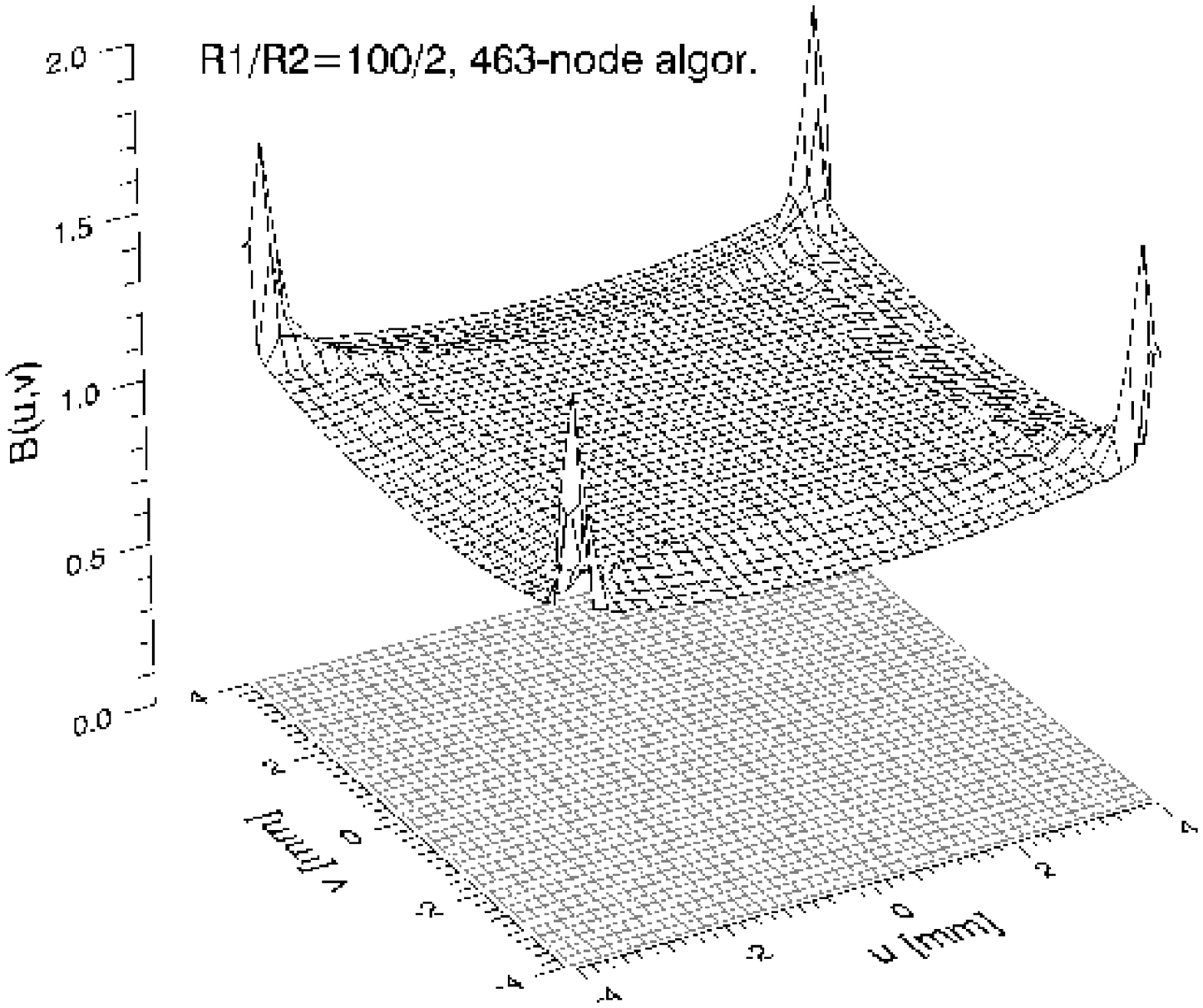}
 \end{center}
  \caption{
            Population density $\widetilde{B}(u,v)$ plotted for the 463-node algorithm
            for $R1=100\,\mathrm{k\Omega}/\Box$ and
            $R2=2\,\mathrm{k\Omega}/\Box$ obtained by
            Eq. (\ref{eq_population_density}) with a homogeneous
            illumination $B(x,y)=1$.
          }
 \label{fig_PopDensity463er}
\end{figure}

A graphical illustration of the matrices $A_x$ and $B_x$ are shown
in Figs. \ref{fig_matrix_a} and \ref{fig_matrix_b}.
The 4-node algorithm determines already the optimum positions in
$x$-direction close to the low resistivity cell borders (except
close to the readout nodes), therefore it is the only algorithm
applied ($a_x=1$). In $y$-direction close to the cell borders the
4-node algorithm becomes unfavourable (cf. Fig.
\ref{fig_4er_6er_3er_Algo} \textbf{a)}) and the other
participating algorithms are more emphasised ($a_x\approx 0$). As
expected the 6-node algorithm offers the best spatial resolution
close to the symmetry axis of the cell, whereas the 3-node
algorithm becomes more favourable in $x$-direction close to the
low resistivity cell borders and the readout nodes (cf. also Fig.
\ref{fig_4_6_3SpatialResUncorrected}). Fig. \ref{fig_463node-algo}
shows the reconstructed positions obtained with the 463-node
algorithm by means of Eq. (\ref{eq_463-node-algo}).
The slight distortions for some positions mainly close to the
nodes are due to the constraint in Eq.
(\ref{eq_constraint_Matrix_a}) and can be directly revealed in the
population density plot (Fig. \ref{fig_PopDensity463er}). A very
slight inhomogeneity is also visible close to the low resistivity
strips at the cell borders. In comparison with the population
density of the 4-node algorithm (Fig. \ref{fig_PopDensity4er}) the
population density of the 463-node algorithm is nearly flat and
especially close to the cell borders and the readout nodes
improved by a factor 2--4.

\begin{figure}
  \vspace{0mm}
  \begin{center}
  \includegraphics[clip,width=7.5cm]{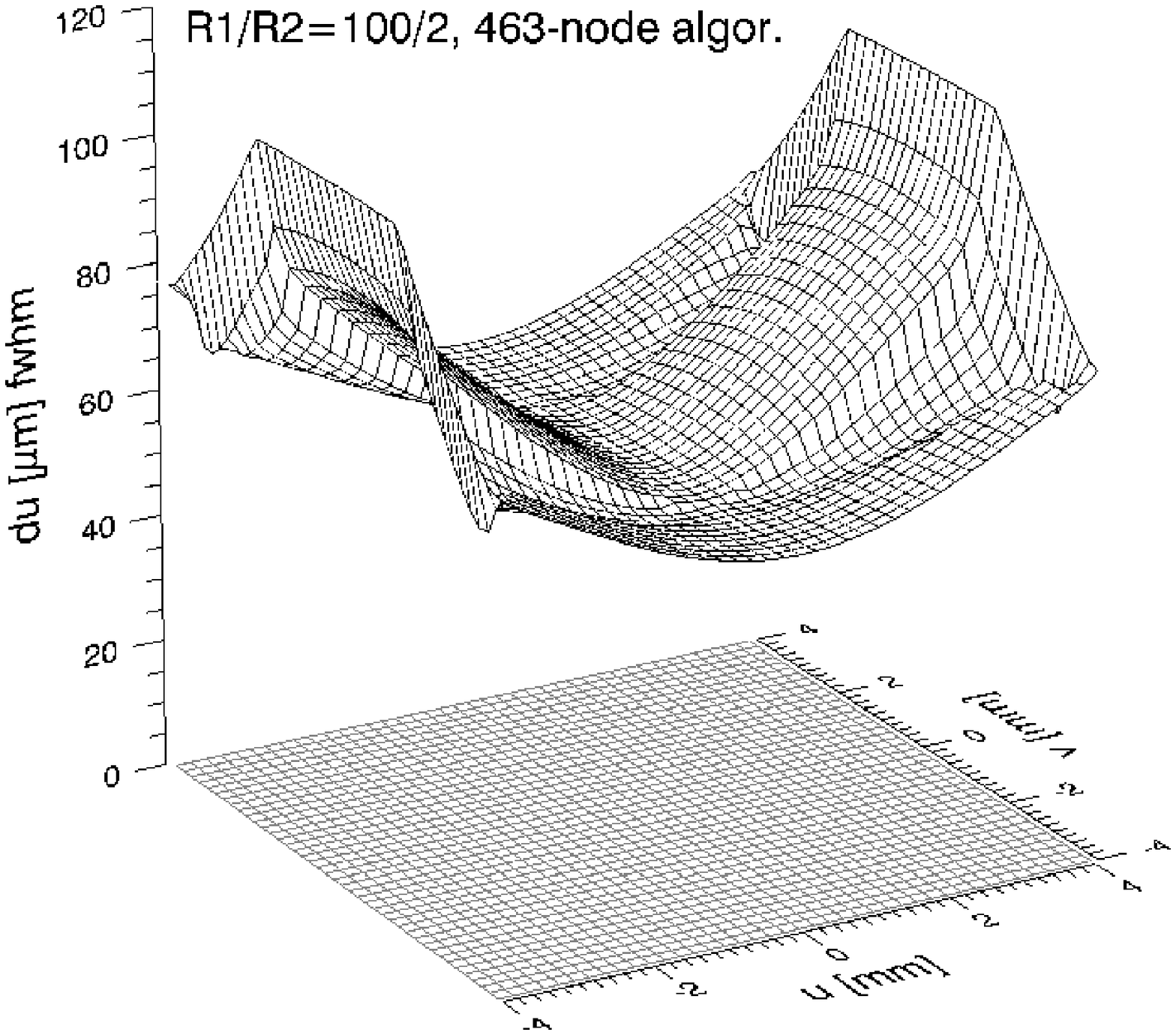}
 \end{center}
  \caption{
            Spatial resolution $du=2.355\cdot\sigma_u$ (fwhm) for
            the 463-node algorithm
            as a function of the reconstructed
            positions $(u,v)$ for
            $R1=100\,k\Omega/\Box$ and $R2=2\,k\Omega/\Box$.
            The graph is plotted for the
            same parameters like in Fig.
            \ref{fig_4_6_3SpatialResUncorrected}.
          }
 \label{fig_463er_SpatialResUncorrected}
\end{figure}

Besides this as a consequence also the spatial resolution is
improved. In Fig. \ref{fig_463er_SpatialResUncorrected} the
spatial resolution of the 463-node algorithm is plotted as a
function of the reconstructed positions $(u,v)$.
In fact the error propagation (Eq. (\ref{eq_error_propagation})
and Eq. (\ref{eq_463-node-algo})) leads for the 463-node algorithm
to a non-trivial equation for the spatial resolution. However, as
expected due to the mixing matrices $A_x$ and $B_x$ close to the
nodes and the cell borders in $v$-direction the spatial resolution
is comparable to that obtained by the single 3-node and 6-node
algorithm (cf. Fig. \ref{fig_4_6_3SpatialResUncorrected}). Close
to the cell borders in $u$-direction the behaviour of the single
4-node algorithm is reflected. Since the positions are indeed well
reconstructed (Fig. \ref{fig_463node-algo}) the spatial resolution
of the 463-node algorithm as a function of the true positions
$(x,y)$ is very similar to that of the reconstructed positions
$(u,v)$ (Fig. \ref{fig_463er_SpatialResUncorrected}). Only close
to the nodes the spatial resolution is slightly deteriorated
caused by the small position corrections (s. Eq.
(\ref{eq_SpatialResTrafo})).

\subsection{Application of the 463-node algorithm on measured data}

The application of the 463-node algorithm (Eq.
\ref{eq_463-node-algo}) demands for the mixing matrices $A$ and
$B$. Since these matrices are actually a function of the true
positions $(x,y)$ which are not known a priori in a measurement we
have to apply an iterative technique. To obtain a first
approximate position an assumption for suitable elements $a_{ij}$
and $b_{ij}$ is needed. For that reason the reconstructed position
is calculated firstly both by means of the $6$-node and the 4-node
algorithm. At a distance of $d=0.1\,g$ (with $g$ as the cell size)
around the cell borders the $6$-node algorithm $(u_6,v_6)$ is used
for a first estimation of $i,j$ to obtain the matrix elements
$a_{ij}$ and $b_{ij}$. Else the 4-node algorithm $(u_4,v_4)$ is
used for the estimation of $a_{ij}$ and $b_{ij}$. Then, the
463-node algorithm position is calculated for the first time,
which gives new positions and therefore an estimation of new
$a_{ij}$ and $b_{ij}$. This iterative procedure is repeated
several times until the reconstructed positions
$(u_{463},v_{463})$ converge. Since the diffusion simulation gives
matrices $A$ and $B$ with $41\times 41$ elements corresponding to
a grid spacing of $200\,\mathrm{\mu m}$ (Figs. \ref{fig_matrix_a}
and \ref{fig_matrix_b}) we have used an interpolation routine of
IDL \cite{IDL} to achieve a finer resolution. If the virtual pixel
size is e.g. chosen to $200\times 200 \,\mathrm{\mu m^2}$ we use
finer matrices $A$ and $B$ with $81\times 81$ elements,
corresponding to $100\,\mathrm{\mu m}$ grid spacing to allow a
better convergence of the 463-node algorithm. Since this method
converges at almost all positions within the cell this iterative
loop is used for all image reconstructions of measured data
presented in the following.

\begin{figure}
  \vspace{0mm}
  \begin{center}
    \includegraphics[clip,width=7.5cm]{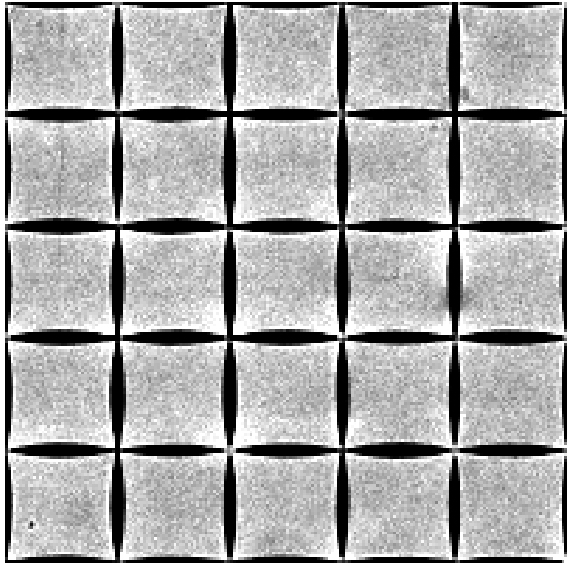}
 \end{center}
  \caption{
            Reconstructed image of a flatfield illumination using
            the 4-node algorithm. A virtual pixel size of
            $200\times 200\,\mathrm{\mu m^2}$ is chosen. The
            image contains about $4.57\cdot 10^6$ photons
            corresponding to a mean number of photons per pixel
            $N\approx 114$. The standard deviation of the intensity
            per pixel distribution amounts to 54.9.
           }
 \label{fig_M4erPopDens}
\end{figure}

Fig. \ref{fig_M4erPopDens} shows the response of the inner
$5\times 5$ cells ($\widehat{=}\,40\times 40\,\mathrm{mm^2}$) of a
detector with a PCB-readout structure to an uniform illumination
($B(x,y)=1$) reconstructed with the 4-node algorithm (Eq.
\ref{eq_4-node-algo}).
A $^{55}\text{Fe}$-source has been used for illumination
($E_{\gamma}=5.9\,\mathrm{keV}$). The detector has been operated
with an $\text{Ar/CO}_2$ (90/10) gas filling at standard pressure.

\begin{figure}
  \vspace{0mm}
  \begin{center}
    \includegraphics[clip,width=7.5cm]{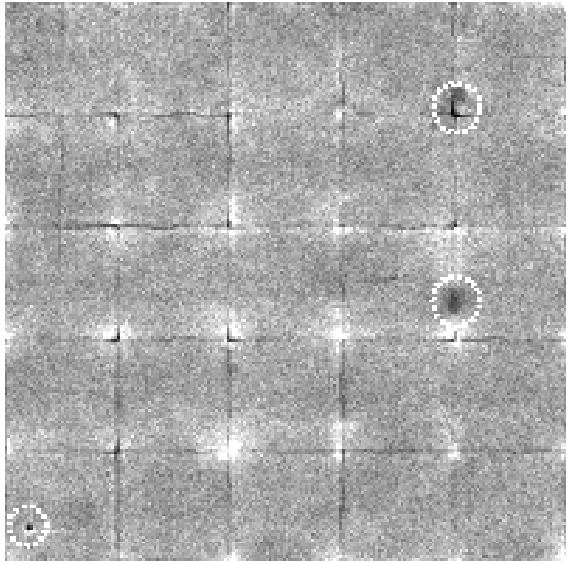}
 \end{center}
  \caption{
            Reconstructed image from the same data as in Fig. \ref{fig_M4erPopDens}
            using the 463-node algorithm. A virtual pixel size of
            $200\times 200\,\mathrm{\mu m^2}$ is chosen.
            The standard deviation of the intensity
            per pixel distribution amounts to 16.2.
            The dotted circles indicate systematic effects.
            \emph{Uppermost circle:} not properly working
            preamplifier.
            \emph{Middle circle:} low resistivity strip is interrupted.
            \emph{Lowermost circle:} defect in a copper layer
            in one of the three GEM structures used for gas amplification.}
 \label{fig_M463erPopDens}
\end{figure}

The low resistivity cell borders of the PCB-structure have a width
of about $(170\pm5\,\mathrm{\mu m})$. The special resistive
materials used for the silk-screen printing process have surface
resistances of $R1=100\,\mathrm{k\Omega}/\Box$ and
$R2=1\,\mathrm{k\Omega}/\Box$, respectively. Unfortunately the
printing and burning processes have a strong influence on the
resistances and the exact ratio of $R1/R2$ is hardly predictable.
Since the depletions at the cell borders when using the 4-node
algorithm show a good regularity in all cells (Fig.
\ref{fig_M4erPopDens}) we conclude that the relative accuracy of
the surface resistances is in the percent range. The resulting
image when using the 463-node algorithm is shown in Fig.
\ref{fig_M463erPopDens}.
The matrices $A$ and $B$ are created for
$R1=100\,\mathrm{k\Omega}/\Box$ and $R2=10\,\mathrm{k\Omega}/\Box$
since this ratio simply fits best, whereby the distortions of the
4-node algorithm depend less than linear on the ratio $R1/R2$.
This means that one obtains e.g. only little changes concerning
the distortions when treating a ratio of $R1/R2=10$ instead of
$R1/R2=20$. The population density $B(u_{463},v_{463})$ becomes
much more homogeneous in comparison to the 4-node algorithm. In
addition the standard deviation of the intensity per pixel is
dramatically reduced from $\sigma_N=54.9$ to $\sigma_N=16.2$ which
is indeed much closer to the theoretical limit of
$\sigma_{N-\text{Poisson}}=\sqrt{114}\approx 10.7$.

\section{Non-linear corrections
\label{Sec_NonLinCorrections}}

Although the reconstructed image with the 463-node algorithm (Fig.
\ref{fig_M463erPopDens}) represents a mentionable progress
compared to images reconstructed e.g. with the 4-node algorithm
(Fig. \ref{fig_M4erPopDens}) still some inhomogeneities in the
population density are obvious. These artifacts can mainly be
attributed to systematic effects (which partly occur during the
measurement). As already described, even in the theoretical case
the 463-node algorithm leads to a slight overpopulation close to
the cell borders and especially around the nodes (cf. Fig.
\ref{fig_PopDensity463er}). Systematic electronic effects, like
gain variations of the preamplifiers and cross-talk in the
preamplifiers/cables can possibly influence the image. Besides
this, a too short integration time of the preamplifiers can lead
to an overpopulation (as can be shown by simulations) of the cell
borders. Further systematic effects stem from the readout
structure itself, like the finite dimensions of the readout nodes,
the finite accuracy of the width of the low resistivity strips and
local inhomogeneities of the surface resistances.

\subsection{Homogeneous population density}

\begin{figure}
  \vspace{0mm}
  \begin{center}
    \includegraphics[clip,width=7.5cm]{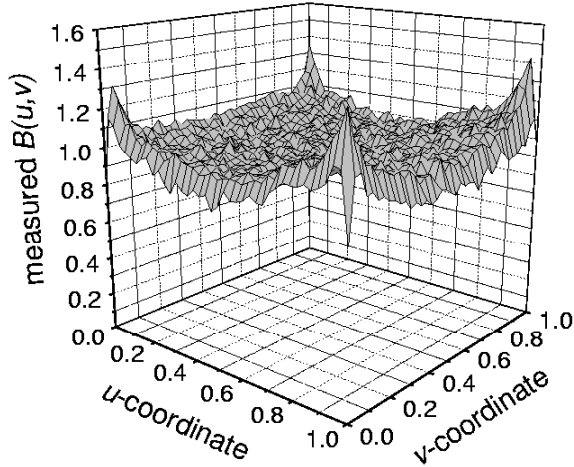}
  \end{center}
  \caption{
            Measured averaged population density $\widetilde{B}(u,v)$ of the $5\times 5$ cells
            of Fig. \ref{fig_M463erPopDens}. The average value of
            the population density is normalised to unity.
            }
 \label{fig_GlobalPopDens463}
\end{figure}

\begin{figure}
  \vspace{0mm}
  \begin{center}
    \includegraphics[clip,width=7.5cm]{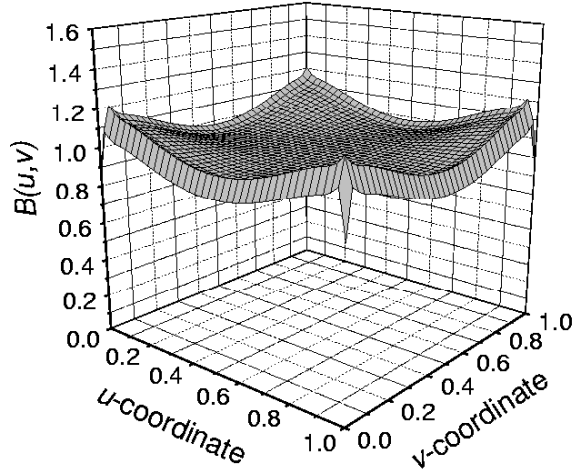}
 \end{center}
  \caption{
            Population density $\widetilde{B}(u,v)=\partial_u x\,\partial_v
            y-\partial_v x\,\partial_u y$ obtained by the
            coordinate transformation functions $x(u,v)$ and $y(u,v)$.
          }
 \label{fig_NonLinPopDensTheo}
\end{figure}

\begin{figure}
  \vspace{0mm}
  \begin{center}
    \includegraphics[clip,width=7.5cm]{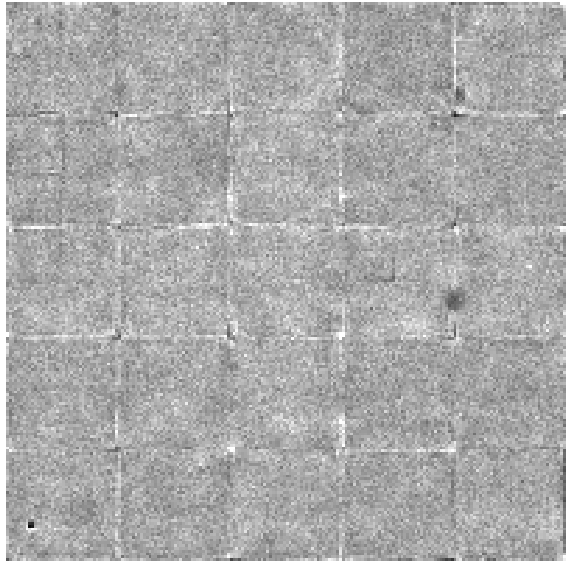}
 \end{center}
  \caption{
            Reconstructed image of the flatfield illumination using
            the 463-node algorithm with subsequent non-linear
            corrections.
            The standard deviation of the intensity
            per pixel distribution amounts to 13.6.
           }
 \label{fig_M463erPopDensNonLin}
\end{figure}

In order to decrease the remaining inhomogeneities around the
readout nodes and the cell borders (cf. Fig.
\ref{fig_M463erPopDens}) an ansatz is made for $x(u,v)$ and
$y(u,v)$, whereby the abbreviation $u=u_{463}$ and $v=v_{463}$ is
used in the following. Since we know that the incident population
density $B(x,y)$ for a homogeneous illumination is equal to unity
Eq. (\ref{eq_population_density}) reads as follows:
\begin{align}
    B(x,y)=1=\frac{\widetilde{B}(u,v)}{\displaystyle\frac{\partial x}{\partial u}\frac{\partial y}{\partial v}
                                     -\frac{\partial x}{\partial v}\frac{\partial y}{\partial
                                     u}}
    \quad\Longleftrightarrow\quad
    \widetilde{B}(u,v) = \frac{\partial x}{\partial u}\frac{\partial y}{\partial v}
                        -\frac{\partial x}{\partial v}\frac{\partial y}{\partial u}
\end{align}
The functions $x(u,v)$ and $y(u,v)$ have now to be optimised in
such a fashion that the derivatives $\partial_u x\,\partial_v
y-\partial_v x\,\partial_u y$ correspond to the measured
population density $\widetilde{B}(u,v)$. Fig.
\ref{fig_GlobalPopDens463} shows the measured mean population
density $\widetilde{B}(u,v)$ obtained by the superposition of the
$5\times 5$ cells shown in Fig. \ref{fig_M463erPopDens}.
The mean value of $\widetilde{B}(u,v)$ is normalised to unity;
also the cell size $g$ is normalised to one ($u,v\in[0,1]$). We
have composed the coordinate transformation functions $x(u,v)$ and
$y(u,v)$ by suitable combinations of different one- and
two-dimensional Gaussian functions with the cell centre as the
point of symmetry. The final coordinate transformation functions
$x(u,v)$ and $y(u,v)$ have altogether 6 independent parameters.
Fig. \ref{fig_NonLinPopDensTheo} shows the population density as a
result of the minimisation of $|\widetilde{B}(u,v)-(\partial_u
x\,\partial_v y-\partial_v x\,\partial_u y)|^2$.
The measured population density $\widetilde{B}(u,v)$ (Fig.
\ref{fig_GlobalPopDens463}) and the population density obtained by
the coordinate transformation functions $x(u,v)$ and $y(u,v)$
(Fig. \ref{fig_NonLinPopDensTheo}) are in good agreement. The
knowledge of $x(u,v)$ and $y(u,v)$ allows now to correct for the
small inhomogeneities around the nodes and the cell borders. By
this procedure, the reconstructed positions of the 463-node
algorithm $(u,v)$ are transformed into the corrected positions
$x(u,v)$ and $y(u,v)$. Fig. \ref{fig_M463erPopDensNonLin} shows
the resulting non-linear corrected image.
The standard deviation of the intensity per pixel is further
reduced from $\sigma_N=16.2$ (Fig. \ref{fig_M463erPopDens}) to
$\sigma_N=13.6$ which is even closer to the theoretical limit of
$\sigma_{N-\text{Poisson}}=\sqrt{114}\approx 10.7$.

\subsection{Inhomogeneous population density}

Since all assumptions made in the previous sections are based on a
homogeneous population density $B(x,y)=1$ we have to test the
463-node algorithm (Sec. \ref{Sec_463-node algorithm}) and the
corresponding non-linear corrections separately with an
inhomogeneous population density. This can e.g. be performed with
a suitable collimator. More images (even time-resolved
measurements) in the field of biological and chemical diffraction
samples recorded with the ViP-detector prototype can be found in
Ref. \cite{Orthen2003c}.

\begin{figure}
  \vspace{0mm}
  \begin{center}
 \includegraphics[clip,width=14cm]{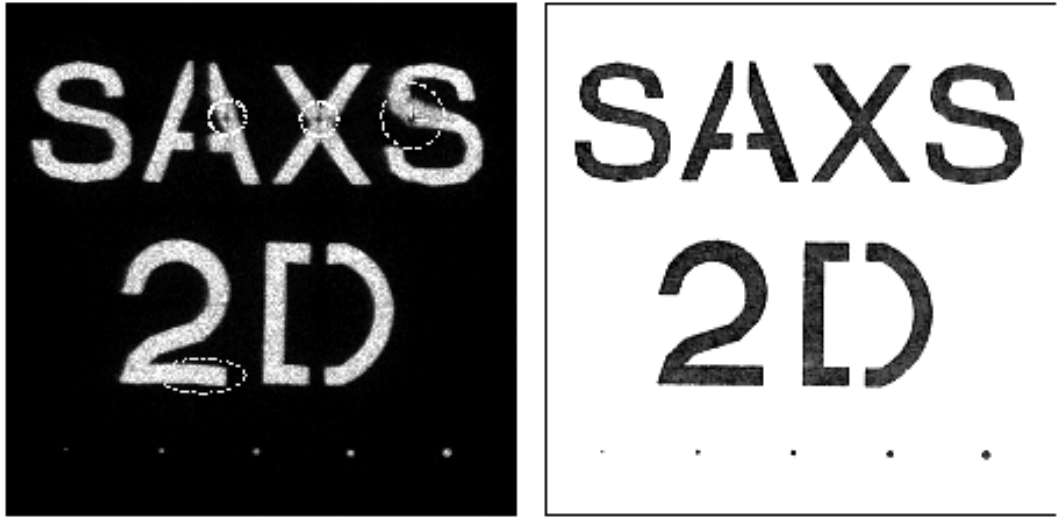}
 \end{center}
  \caption{On the left hand side the image of the ``SAXS" aperture
  recorded with a PCB-readout structure is shown. On the right
  hand side a photo of the aperture taken with a standard digital camera
  against the sunlight is depicted. The distance between detector
  and source amounted to about
  $2.5\,\mathrm{m}$, therefore parallax is very small.
  The illuminated spots at the
  bottom correspond to holes in the aperture with diameters of
  280, 380, 480, 580 and $680\,\mathrm{\mu m}$. The aperture is
  slightly tilted by an angle of about $0.6^{\circ}$, which can
  be recognised for example by the vertical pixel jump (pixel quantisation)
  at the bottom of the ``2" (dashed-dotted line). The sizes of
  both depicted images amount to $4.4\times4.4\,\mathrm{cm}^2$.
  }
 \label{fig_SAXScollimator}
\end{figure}

Fig. \ref{fig_SAXScollimator} shows the image of a laser cut
$1\,\mathrm{mm}$ thick stainless steel aperture containing ``SAXS
2D" letters (\textbf{S}mall \textbf{A}ngle \textbf{X}-ray
\textbf{S}cattering) and five holes with increasing hole
diameters.
The image has been recorded with a PCB-readout structure using
photons of an energy of $6.4\,\mathrm{keV}$ (fluorescence of a
Fe-target in a $8\,\mathrm{keV}$ synchrotron beam). The detector
was filled with a $\text{Xe/CO}_2$ (90/10) gas mixture at a
pressure of $1.3\,\mathrm{bar}$. The image has been reconstructed
with the linear 463-node algorithm with subsequent non-linear
corrections at the nodes and at the cell borders. We performed an
additional flatfield correction with a flatfield image recorded in
our laboratory one week after the aperture measurement with a
different gas filling and a different X-ray source. Since this
flatfield image is different from one which we would have obtained
at the beamline directly after the aperture measurement (with
identical detector parameters), the corrected image still shows up
some artifacts like visible areas around the readout nodes (marked
with dotted lines) which should have been suppressed by the
correction of the proper flatfield image.

However, the recorded image compares well to the image of the
aperture (right hand side image in Fig. \ref{fig_SAXScollimator}).
Only the middle part of the second ``S" (dashed line) looks
slightly distorted; at this readout channel the preamplifier was
not working optimally (compare to uppermost dotted circle in Fig.
\ref{fig_M463erPopDens}). Also the reproduction of the holes shows
a good agreement.

\section{Conclusion}

The simplest linear algorithms do not reconstruct the event
positions correctly. This leads to distortions since the true
positions $(x,y)$ and the reconstructed positions $(u,v)$ differ
up to a certain extent. When these algorithms are corrected
afterwards by a suitable coordinate transformation connecting the
two spaces $(u,v)$ and $(x,y)$ the spatial resolution becomes
worse especially next to the cell borders and the readout nodes.

Therefore, the 463-node algorithm is introduced.  This complex
algorithm consisting of combinations of linear algorithms is a
possible solution for position reconstructions using the
two-dimensional interpolating resistive readout structure
described above. This algorithm has been optimised with respect to
both the correct position reconstruction and the spatial
resolution. Indeed, the reconstruction of a homogeneous population
density as a measure for the quality of an image has been
considerably improved -- even close to the Poisson limit -- when
the 463-node algorithm with subsequent non-linear corrections was
applied. Nevertheless, some inhomogeneities mainly caused by
systematic effects are still visible in the images. Since these
disturbing effects can not be corrected afterwards they should be
kept as small as possible.

\bibliographystyle{elsart-num}
\bibliography{reflist}

\end{document}